\PassOptionsToPackage{table}{xcolor}
\documentclass[sigconf,nonacm,balance=false]{acmart}
\setcopyright{none}
\hypersetup{keeppdfinfo=true}
\usepackage{amsmath}
\usepackage{algorithm}
\usepackage{algorithmic}
\usepackage{newfloat}
\usepackage{multirow}
\usepackage{listings}
\usepackage{colortbl}
\usepackage{tabularx}
\usepackage{placeins}
\DeclareCaptionStyle{ruled}{labelfont=normalfont,labelsep=colon,strut=off}
\floatstyle{ruled}
\newfloat{listing}{tb}{lst}{}
\floatname{listing}{Listing}
\ifdefined\pdfrunninglinkoff
  \AddToHook{fancyhdr/before}{\pdfrunninglinkoff}
  \AddToHook{fancyhdr/after}{\pdfrunninglinkon}
\fi

\AtBeginDocument{\raggedbottom}
\begin{document}

\title[Search When Needed, Proceed When Reliable]{Search When Needed, Proceed When Reliable: Metacognitive Reliability Supervision for Efficient Agentic Search}

\author{Minhao Sun}
\affiliation{\institution{Beijing Institute of Technology}
  \city{Beijing}
  \country{China}
}

\author{Zhijing Wu}
\authornote{Corresponding author: \href{mailto:zhijingwu@bit.edu.cn}{zhijingwu@bit.edu.cn}.}
\affiliation{\institution{Beijing Institute of Technology}
  \city{Beijing}
  \country{China}
}

\author{Xian-Ling Mao}
\affiliation{\institution{Beijing Institute of Technology}
  \city{Beijing}
  \country{China}
}
\renewcommand{\shortauthors}{Sun, Wu, and Mao}

\begin{abstract}
Efficient agentic search requires large language models (LLMs) to determine whether they can reliably proceed with their current knowledge or need additional external evidence. Supervision based on final-answer correctness or trajectory-level feedback evaluates realized trajectory performance. However, observed trajectory performance is jointly shaped by stochastic generation, retrieval quality, and downstream reasoning. Such signals therefore provide only indirect evidence of how reliable the current decision state was and how retrieved evidence changed its reliability. An effective reward should encourage search when external information is needed and favor searches that contribute to the answer. To this end, we propose a metacognitive framework for training LLMs to make search decisions based on the reliability of their intermediate reasoning states. Our framework estimates the reliability of local conclusions to distinguish whether the current information supports proceeding without search from how much retrieved evidence improves that reliability. The current-state reliability score assesses whether the model can proceed without search. The search-action score measures how retrieved evidence changes estimated reliability. Together, these scores define a reliability-aware process reward for Group Relative Policy Optimization (GRPO), providing finer-grained supervision for search and no-search behavior. Experiments across seven question-answering benchmarks demonstrate that our method achieves a better accuracy--efficiency trade-off than strong agentic-search baselines. Our code and model checkpoints will be open-sourced.

\end{abstract}


\maketitle
\pagestyle{standardpagestyle}

\section{Introduction}
\begin{figure}[!t]
\centering
\includegraphics[width=1.0\columnwidth,trim=0 72bp 0 0,clip]{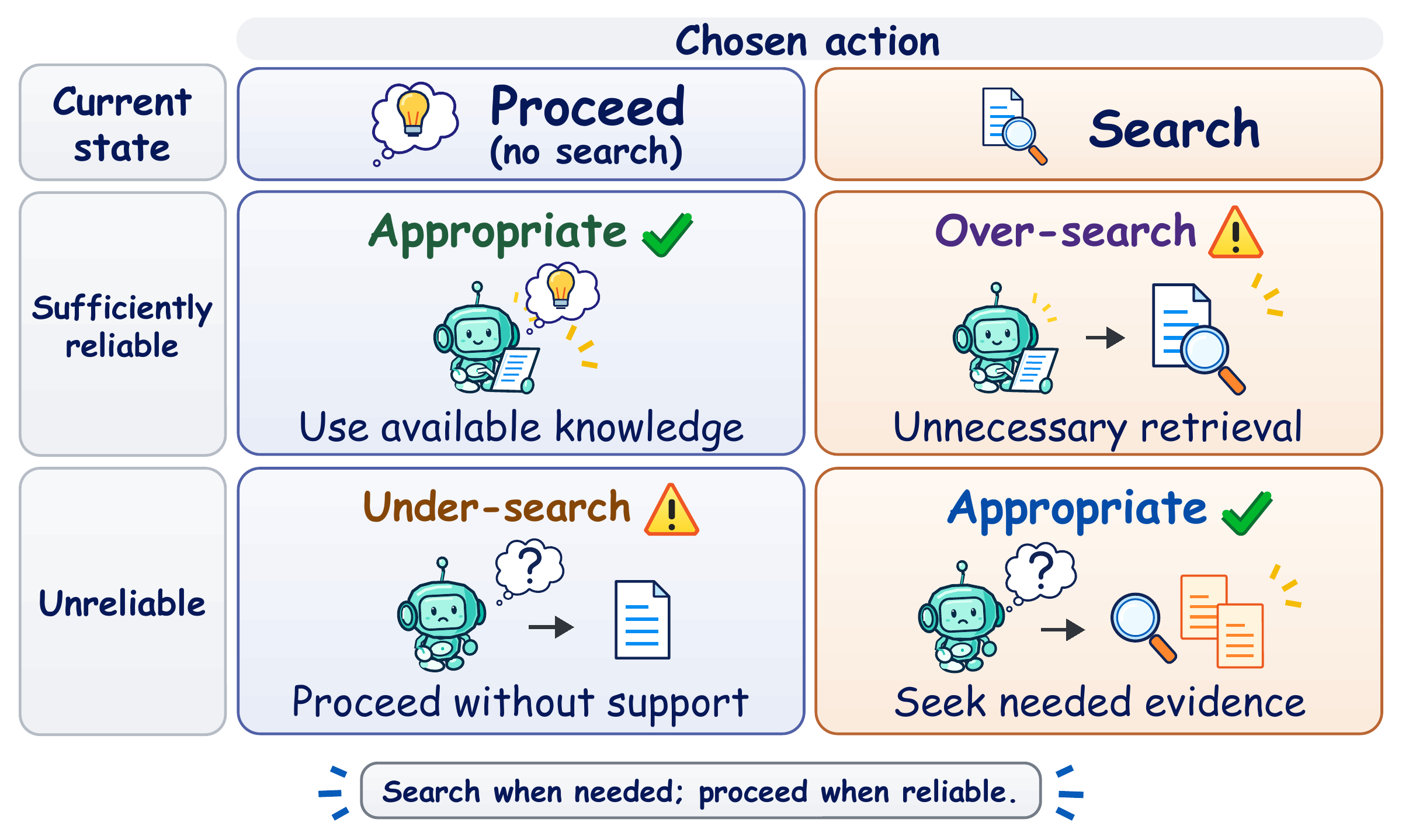}
\caption{Conceptual decision matrix for efficient agentic search. Under-search proceeds from an unreliable state; over-search retrieves despite sufficient reliability. The matrix illustrates the reliability-aware search-control objective.}
\label{fig:decision-problem}
\end{figure}

Agentic retrieval-augmented generation (RAG) and search-augmented reasoning support Web information access by allowing large language models (LLMs) to acquire and integrate external evidence over multiple reasoning steps~\cite{trivedi2023ircot,li2025searcho1,jin2025searchr1,mei2026searchart,tao2026race}. This flexibility improves knowledge-intensive question answering, but it also introduces a recurring decision: at each intermediate state, should the model search or proceed using its current parametric knowledge and accumulated evidence? A common way to supervise this choice is to reward trajectories that produce correct answers. Yet a trajectory's final outcome does not always reveal whether its search decisions were appropriate. A correct answer without search may reflect either reliable knowledge or an unsupported lucky guess. Even informative retrieval can be followed by an incorrect final answer because later reasoning fails. A correct trajectory may also contain searches that were never necessary. These cases expose a performance--reliability mismatch. As illustrated in Figure~\ref{fig:decision-problem}, overestimating current reliability can lead to under-search, whereas underestimating it can lead to over-search. The key distinction is between whether a conclusion happened to be correct, how reliably it could be reached from the current information, and how much retrieval improved that reliability. Our process supervision targets the latter two quantities alongside outcome rewards.

To distinguish these states, we estimate how reliably the model can resolve the local factual objective from its current context. This indicates whether the available information supports the next reasoning step. We use sequence uncertainty as a signal for this estimate: Figure~\ref{fig:qwen7b-reliability} shows that local-conclusion accuracy generally decreases with mean sequence uncertainty, consistent with prior observations~\cite{chen2026knowclearer}.

Our metacognitive framework uses these reliability estimates to supervise search and no-search decisions. The \emph{Current-State Reliability Score} evaluates whether the model can proceed without search by comparing its estimated reliability with a calibration reference. The \emph{Search-Action Reliability Score} measures the change between matched contexts with and without retrieved evidence. Averaging these scores within each rollout yields a process reward, which is added to answer and format rewards for GRPO. Together, these signals encourage the policy to proceed when current information is reliable and to favor searches that improve reliability. We summarize our contributions as follows:

\begin{samepage}
\begin{itemize}
\interlinepenalty=10000
\item \textbf{Correctness versus reliability.} We distinguish whether a local conclusion is correct from how reliably it can be reached with the information available at that decision, clarifying why realized correctness alone is an ambiguous signal for search supervision.
\item \textbf{Action-specific reliability scoring.} Using local probes from the current policy, we score no-search decisions by current-state reliability and search decisions by the reliability change between matched contexts with and without retrieved evidence. Averaging these action scores within each rollout yields a process reward that augments answer and format rewards in GRPO.
\item \textbf{Improved accuracy and adaptive search allocation.} Our 7B model achieves the highest average CEM across seven QA benchmarks (69.3), exceeding HiPRAG, the strongest baseline, by 2.1 points and improving on it across all seven datasets, as reported in Table~\ref{tab:hiprag-original-results}. In controlled experiments, our policy reallocates search across task types: relative to our HiPRAG reimplementation, mean search calls decrease by approximately 20.9\% on single-hop QA and increase by 6.6\% on multi-hop QA, with 6.8\% fewer searches overall, as shown in Table~\ref{tab:search-calls-selected}.
\end{itemize}
\end{samepage}

\section{Related Work}
\subsection{Agentic RAG and Search-Augmented Reasoning}

Retrieval-augmented generation conditions language models on external evidence for knowledge-intensive tasks~\cite{lewis2020retrieval}. IRCoT interleaves retrieval with chain-of-thought reasoning~\cite{trivedi2023ircot}, allowing later queries to depend on information acquired earlier in the reasoning chain. Search-o1 incorporates search into extended reasoning~\cite{li2025searcho1}, while DRAGIN estimates information needs during generation to determine when and what to retrieve~\cite{su2024dragin}. Search-R1 and R1-Searcher learn to interleave reasoning and retrieval through reinforcement learning with outcome rewards~\cite{jin2025searchr1,song2025r1searcher}; Smart-Searcher further encourages selective knowledge acquisition~\cite{song-etal-2025-smart-searcher}. These methods make retrieval a decision within reasoning, requiring the policy to determine when its available information is sufficient and when additional evidence is needed.

\subsection{Efficient Search Control and Process Rewards}

Search-control objectives account for the conditions under which an answer is obtained. SEM rewards correct direct answers and successful search-assisted revision of incorrect preliminary answers~\cite{sha2025sem}, while SAAS compares search-enabled and search-disabled rollouts to adapt search penalties to the model's knowledge boundary~\cite{tang2026saas}. DAS uses interventions on search and answer actions to construct trajectory preferences from their effects on correctness~\cite{zhang2026tosearch}. BAPO addresses answer reliability by adaptively rewarding explicit uncertainty responses when no sampled rollout answers correctly~\cite{liu2026bapo}. Search Wisely relates inefficient retrieval to uncertainty about the model's knowledge boundary and encourages confident search decisions that lead to correct answers~\cite{wu-etal-2025-search-wisely}.

HiPRAG gates its process reward by answer and format correctness and evaluates steps through retrieval redundancy and local correctness~\cite{wu2026hiprag}. Under this criterion, a search is treated as redundant when its conclusion can also be obtained without retrieval.

Our process supervision assesses how reliably current information supports a local conclusion and how much retrieved evidence improves that reliability. The resulting process reward is added independently of final-answer and format correctness. A correct conclusion can arise from an unreliable state, and evidence can improve estimated reliability without changing the conclusion's content.

\subsection{Uncertainty and Calibration}

Answer reliability can be estimated from model-derived uncertainty. Single-sequence likelihood measures use model probabilities to estimate predictive uncertainty~\cite{measurerethinking}, while MIND uses internal representations for hallucination detection without manual annotations~\cite{su-etal-2024-unsupervised}. For sequence uncertainty, empirical mappings relate the measured value to observed answer accuracy. Know More, Know Clearer estimates the relationship between sequence uncertainty and answer accuracy, then distinguishes mastered, confused, and missing knowledge to guide targeted knowledge augmentation and motivate confidence calibration~\cite{chen2026knowclearer}.

We turn this relationship into action-specific search supervision by probing the reliability of intermediate conclusions under matched contexts with and without retrieved evidence.

\section{Preliminaries and Motivation}

\subsection{Sequence-level Uncertainty}
For a local conclusion $y=(x_1,x_2,\ldots,x_T)$ generated from context $c$, sequence-level uncertainty is the mean negative log-likelihood of its generated tokens~\cite{measurerethinking}:
\[
U(y\mid c)=-\frac{1}{T}\sum_{t=1}^{T}\log p_{\theta}(x_t\mid c,x_{<t}).
\]
The context contains the question and local reasoning state; only conclusion tokens enter the average.

Lower uncertainty indicates that the conclusion tokens are more predictable given the context. We then empirically investigate whether this greater predictability is associated with higher accuracy.

\subsection{Empirical Uncertainty–Accuracy Relationship}
\begin{figure}[!t]
\centering
\includegraphics[width=0.92\linewidth]{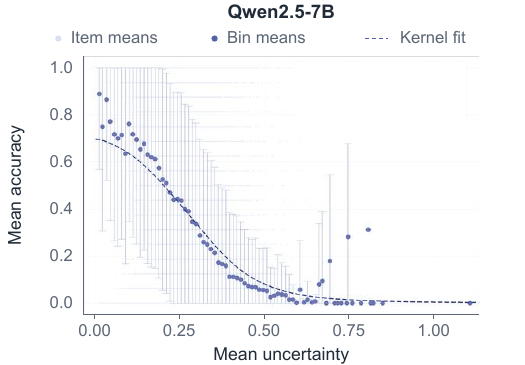}
\caption{Local uncertainty--accuracy relationship for Qwen2.5-7B: 50,000 states, 32 conclusions per state. Bars indicate within-bin SD.}
\label{fig:qwen7b-reliability}
\end{figure}

Figure~\ref{fig:qwen7b-reliability} shows that local accuracy decreases with mean conclusion uncertainty for Qwen2.5-7B. We sample 32 conclusions per search state without the current retrieval evidence. Let $\overline{U}$ and $\overline{A}$ denote state-level mean uncertainty and local accuracy. For backbone $m$,

\[
\mathbb{E}[\overline{A}\mid \overline{U},m] \approx f_m(\overline{U}).
\]

This relationship provides an empirical basis for estimating the reliability of local conclusions. For search supervision, we distinguish between the reliability supported by the current information and the change in reliability associated with retrieved evidence.

\section{Methodology}

\begin{figure*}[!t]
\centering
\includegraphics[width=\textwidth]{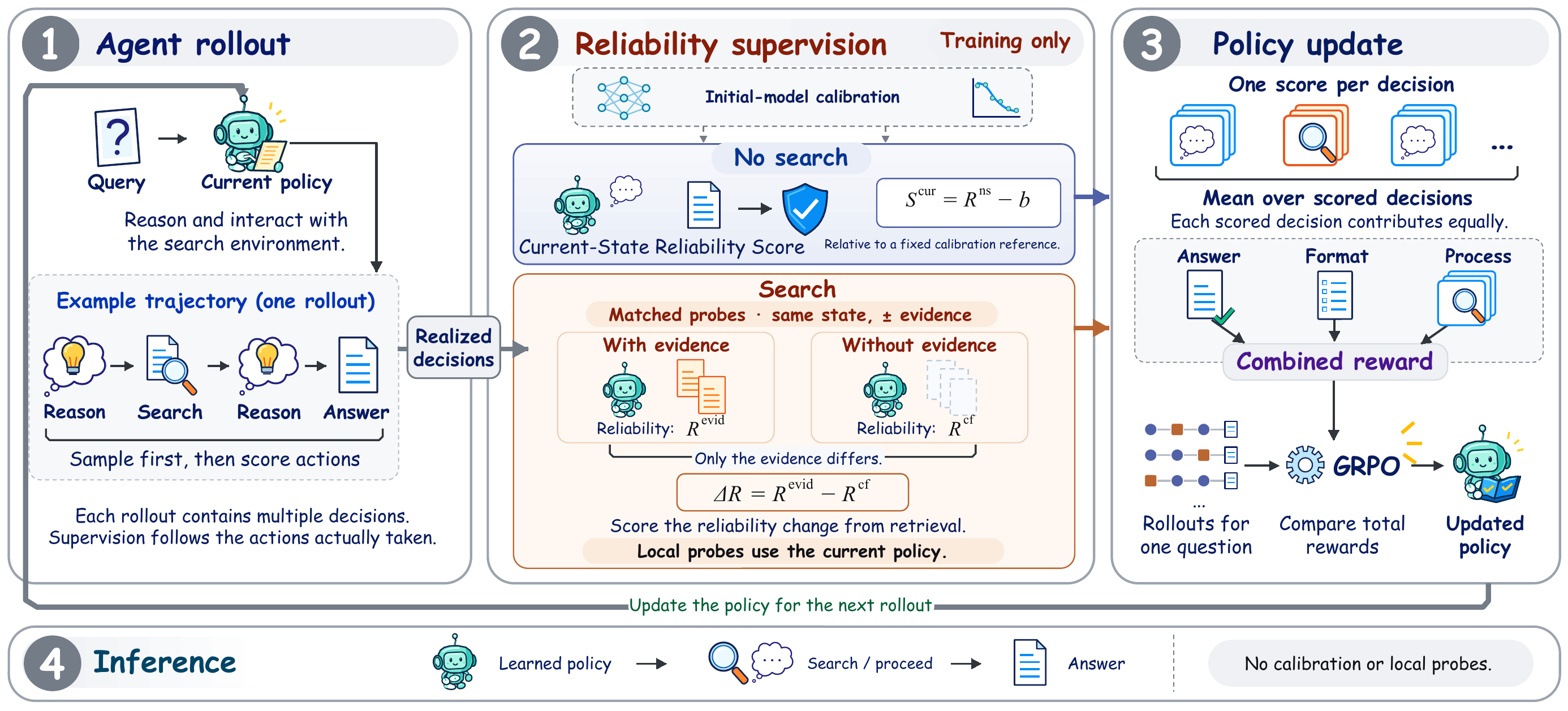}
\caption{Reliability-based process supervision. Local probes estimate reliability for scoring each action; the rollout mean of these scores augments answer and format rewards for GRPO.}
\label{fig:overview}
\end{figure*}

Given a question, the policy generates a sequence of reasoning steps and a final answer. Each step ends with a local conclusion: a factual statement or intermediate answer derived from the current reasoning context and available evidence. A local conclusion can resolve a subquestion that supports later reasoning. In a no-search step, the policy proceeds from its existing information; in a search step, it issues a query and incorporates the returned passages before concluding. As illustrated in Figure~\ref{fig:overview}, our process reward evaluates the reliability of these local conclusions and the change induced by retrieval, assigning supervision to the search decisions within each rollout.

\subsection{Estimating Local Reliability}
\label{sec:reliability-mapping}
\label{sec:local-probing}

The mapping provides a common empirical reliability scale; the local probes determine which decision context is evaluated on that scale.

\noindent\textbf{Empirical mapping.} Before training, we fit a mapping $f_m$ from sequence uncertainty to empirical accuracy for each backbone $m$. Each frozen backbone samples 32 local conclusions per state without the current retrieval evidence. We fit a smooth curve to state-level mean uncertainty and judged local accuracy. The mapping takes values in $[0,1]$ and remains fixed during policy optimization.

We divide the observed uncertainty range into 100 equal-width intervals. For interval $j$, let $\Phi_j$, $\Psi_j$, and $n_j$ denote its mean state-level uncertainty, mean local accuracy, and state count. We retain intervals indexed by $\mathcal{J}_{50}=\{j:n_j\geq 50\}$ and estimate reliability as
\begin{equation}
f_m(u)=
\frac{\sum_{j\in\mathcal{J}_{50}}n_j
 [K_h(u-\Phi_j)+K_h(u+\Phi_j)]\Psi_j}
{\sum_{j\in\mathcal{J}_{50}}n_j
 [K_h(u-\Phi_j)+K_h(u+\Phi_j)]},
\label{eq:empirical-mapping}
\end{equation}
where $K_h(d)=\exp[-d^2/(2h^2)]$. The weights $n_j$ account for the number of calibration states in each interval, while the mirrored kernel term handles the nonnegative uncertainty boundary. The bandwidth and calibration protocol are specified in Appendix~\ref{Appendix:ReliabilityMappingConstruction}.

\noindent\textbf{Matched local probes.} For each decision, we remove its original local conclusion and independently sample replacements from the current policy under each context condition. A no-search probe retains the original question, preceding trajectory, and current reasoning. A search probe also retains the query chosen by the policy. The evidence-conditioned probe includes the passages returned by that query; the no-evidence probe omits only those passages. Earlier retrieved information remains in the shared trajectory history. Both search conditions exclude the original conclusion, so the comparison uses the same reasoning and query without carrying the evidence-conditioned conclusion into the no-evidence context.

\noindent\textbf{From probes to reliability.} We compute each probe's uncertainty from its newly generated conclusion tokens using the length-normalized measure in Section~3.1. The trajectory prefix supplies the conditioning context. For each context condition, we average the probe uncertainties to obtain $\overline{U}$ and estimate reliability as
\[
R=f_m(\overline{U}).
\]

Each conclusion contributes equally to the mean, and the mapping is applied after averaging uncertainty. The current policy supplies both the sampled conclusions and their token probabilities.

Search probes reuse the original query and, in the evidence-conditioned case, its returned passages, without additional retrieval calls. Decoding settings and the probing protocol are given in Appendices~\ref{Appendix:Hyperparameters} and~\ref{Appendix:ReliabilityMappingConstruction}.

\subsection{Scoring Search and No-Search Decisions}

Reliability after a search and the value of that search are not the same: a conclusion may already be reliable without retrieval, while helpful evidence may still leave substantial uncertainty. We therefore score no-search decisions by current reliability and search decisions by evidence-induced gains.

\noindent\textbf{No-search decisions.} Given estimated reliability $R^{\mathrm{ns}}$, the Current-State Reliability Score is
\begin{equation}
S^{\mathrm{cur}} = R^{\mathrm{ns}} - b_m.
\end{equation}
The fixed reference is
\[
b_m=f_m\!\left(
\frac{1}{|\mathcal{D}_{\mathrm{cal}}|}
\sum_{i\in\mathcal{D}_{\mathrm{cal}}}\overline{U}_i
\right),
\]
where $\mathcal{D}_{\mathrm{cal}}$ indexes calibration states and $\overline{U}_i$ is the mean uncertainty of the backbone's sampled local conclusions at state $i$. This rewards no-search decisions above the reference and penalizes those below it. The same reference is used for every decision from a given backbone throughout training.

\noindent\textbf{Search decisions.} Let $R^{\mathrm{evid}}$ and $R^{\mathrm{cf}}$ denote the reliability estimates with and without retrieved evidence for the same search decision. The Search-Action Reliability Score is
\begin{equation}
\Delta R=R^{\mathrm{evid}}-R^{\mathrm{cf}}.
\end{equation}
The score rewards evidence that increases estimated reliability and penalizes evidence that decreases it. Because the query is held fixed, this comparison evaluates the gain from the returned passages in the current reasoning context.

Even when both conditions have high reliability, similar estimates yield a near-zero search score. A search can receive a positive score even when evidence-conditioned reliability remains below the no-search reference: the estimate has increased relative to the matched no-evidence condition. Semantically equivalent conclusions can have different uncertainties under the two conditions, producing a reliability change even when their answer content agrees.

\subsection{Learning with Reliability-Based Rewards}

We generate a group of complete rollouts for each training question using the current policy. The policy determines the search and no-search steps, and the retriever supplies evidence for its queries. After generation, we parse each rollout, construct the local probes, and assign a score to each valid decision according to the action taken. The probes supply reward estimates for the original trajectory; its steps and final answer remain the objects of policy optimization.

We first average the scores of the actions taken in each rollout:
\begin{equation}
R_{\mathrm{process}} =
\frac{1}{|\mathcal{T}|}\sum_{t\in\mathcal{T}}S_t,
\label{eq:process-reward}
\end{equation}
where $S_t$ is the no-search or search score at step $t$. The set $\mathcal{T}$ contains parsed steps with nonempty reasoning and conclusions; search steps also require a nonempty query and nonempty retrieved context. We assign zero process reward when no valid step remains.

Each retained decision contributes once with equal weight. The local probes estimate that decision's score and are not counted as additional trajectory steps. The two action types therefore contribute in proportion to their frequency in the rollout. Since reliability estimates and the reference lie in $[0,1]$, action scores and their mean remain in $[-1,1]$, giving trajectories of different lengths the same process-reward scale.

We train with Group Relative Policy Optimization (GRPO)~\cite{shao2024deepseekmath}, combining the process reward with binary answer and format rewards:
\begin{equation}
\begin{aligned}
R_{\mathrm{total}} ={}&
(1-\lambda_f)R_{\mathrm{answer}}
+ \lambda_f R_{\mathrm{format}} \\
&+ \lambda_p R_{\mathrm{process}}.
\end{aligned}
\label{eq:total-reward}
\end{equation}
Here, $R_{\mathrm{answer}},R_{\mathrm{format}}\in\{0,1\}$ indicate final-answer correctness under CEM and valid trajectory format. The process term is added independently of these indicators. Valid local decisions can therefore contribute positive or negative process supervision even when the final answer is incorrect or the trajectory format is invalid. GRPO compares these totals within each question's rollout group to update the policy. At inference, the policy generates reasoning, searches, and the final answer without calibration or local probing.

\begin{table*}[!t]
\centering
\caption{Main CEM results (\%, higher is better) on seven QA benchmarks. All methods use the Qwen2.5-7B backbone. Bold and underlining mark the best and second-best values per column. Avg. denotes the mean weighted by the evaluation sample counts listed in Table~\ref{tab:evaluation-splits}.}
\label{tab:hiprag-original-results}
\begingroup
\fontsize{9}{10.8}\selectfont
\setlength{\tabcolsep}{3.2pt}
\renewcommand{\arraystretch}{1.12}
\setlength{\heavyrulewidth}{0.7pt}
\setlength{\lightrulewidth}{0.35pt}
\setlength{\cmidrulewidth}{0.3pt}
\arrayrulecolor{black!65}
\begin{tabularx}{\textwidth}{l*{8}{>{\raggedleft\arraybackslash}X}}
\toprule
& \multicolumn{3}{c}{\textbf{Single-hop QA}} & \multicolumn{4}{c}{\textbf{Multi-hop QA}} & \\
\cmidrule(lr){2-4}\cmidrule(lr){5-8}
\textbf{Method} & NQ & TriviaQA & PopQA & HotpotQA & 2Wiki & MuSiQue & Bamboogle & \textbf{Avg.} $\uparrow$ \\
\midrule
\rowcolor{blue!5}
\multicolumn{9}{l}{\rule[-2pt]{0pt}{12pt}\textbf{Non-agentic baselines}} \\
Direct Inference & 27.0 & 26.8 & 40.1 & 58.7 & 16.0 & 7.9 & 15.9 & 31.5 \\
Standard RAG~\cite{lewis2020retrieval} & 51.2 & 54.7 & \underline{65.7} & 56.9 & 21.6 & 18.5 & 18.6 & 48.0 \\
\addlinespace[3pt]
\rowcolor{blue!5}
\multicolumn{9}{l}{\rule[-2pt]{0pt}{12pt}\textbf{Prompt-based agentic RAG}} \\
IRCoT~\cite{trivedi2023ircot} & 27.5 & 36.0 & 42.5 & 51.4 & 37.6 & 19.4 & 20.6 & 39.0 \\
Search-o1~\cite{li2025searcho1} & 40.2 & 42.2 & 58.5 & 56.1 & 45.6 & 15.1 & 19.3 & 48.1 \\
\addlinespace[3pt]
\rowcolor{blue!5}
\multicolumn{9}{l}{\rule[-2pt]{0pt}{12pt}\textbf{RL-based agentic RAG}} \\
R1-Searcher~\cite{song2025r1searcher} & 60.0 & 73.0 & 58.2 & 60.4 & 60.3 & 32.9 & 55.8 & 61.2 \\
R1-Searcher++~\cite{song2025r1searcherplus} & 61.0 & 73.5 & 59.0 & \underline{64.2} & 63.2 & 32.3 & \textbf{58.7} & 62.8 \\
Search-R1~\cite{jin2025searchr1} & 61.2 & 73.6 & 56.5 & 54.0 & 63.6 & 24.8 & 48.4 & 60.4 \\
$\beta$-GRPO~\cite{wu-etal-2025-search-wisely} & 65.0 & 75.0 & 60.0 & 53.0 & 66.0 & 24.0 & 52.0 & 62.4 \\
HiPRAG~\cite{wu2026hiprag} & \underline{71.2} & \underline{76.3} & 63.2 & 62.4 & \underline{71.7} & \underline{34.1} & 52.8 & \underline{67.2} \\
\midrule
\textbf{Ours} & \textbf{74.3} & \textbf{77.7} & \textbf{66.8} & \textbf{66.1} & \textbf{71.8} & \textbf{35.2} & \underline{58.4} & \textbf{69.3} \\
\bottomrule
\end{tabularx}
\endgroup
\end{table*}

\begin{figure*}[!t]
\centering
\begin{minipage}[t]{\dimexpr(\textwidth-\columnsep)/2\relax}
\vspace{0pt}
\captionsetup{type=figure}
\centering
\includegraphics[width=\linewidth]{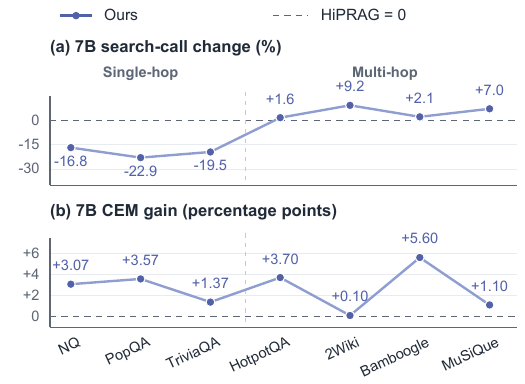}
\caption{Changes relative to HiPRAG (zero baseline). (a) Search-call change (\%) in the controlled 7B evaluation, $100(C_{\mathrm{Ours}}/C_{\mathrm{HiPRAG}}-1)$, using Table~\ref{tab:search-calls-selected}. (b) CEM gain of Ours (7B) over HiPRAG (7B), in percentage points.}
\label{fig:adaptive-search-allocation}
\end{minipage}\hfill
\begin{minipage}[t]{\dimexpr(\textwidth-\columnsep)/2\relax}
\vspace{0pt}
\captionsetup{type=table}
\caption{Mean search calls per question by task type. $\Delta$ is the relative change from HiPRAG. Group and overall means are weighted by evaluation sample counts.}
\label{tab:search-calls-selected}
\begingroup
\fontsize{9}{10.8}\selectfont
\setlength{\tabcolsep}{3pt}
\renewcommand{\arraystretch}{1.18}
\begin{tabularx}{\linewidth}{@{}>{\raggedright\arraybackslash}Xrrr@{}}
\toprule
\textbf{Dataset/group} & HiPRAG & Ours & $\Delta$ calls (\%) \\
\midrule
\rowcolor{blue!5}
\textbf{Single-hop avg.} & \textbf{1.51} & \textbf{1.19} & $\mathbf{-20.9}$ \\
NQ & 1.31 & 1.09 & $-16.8$ \\
PopQA & 1.57 & 1.21 & $-22.9$ \\
TriviaQA & 1.49 & 1.20 & $-19.5$ \\
\midrule
\rowcolor{blue!5}
\textbf{Multi-hop avg.} & \textbf{2.09} & \textbf{2.23} & $\mathbf{+6.6}$ \\
2Wiki & 2.18 & 2.38 & $+9.2$ \\
Bamboogle & 1.90 & 1.94 & $+2.1$ \\
HotpotQA & 1.88 & 1.91 & $+1.6$ \\
MuSiQue & 2.27 & 2.43 & $+7.0$ \\
\midrule
\textbf{Overall} & \textbf{1.76} & \textbf{1.64} & $\mathbf{-6.8}$ \\
\bottomrule
\end{tabularx}
\endgroup
\end{minipage}
\end{figure*}

\section{Experiment Setup}
\subsection{Datasets and Evaluation Metrics}

\paragraph{\upshape\bfseries Training and calibration data}
Following prior work~\cite{wu2026hiprag}, we use the official training sets of Natural Questions (NQ) and HotpotQA for policy optimization~\cite{kwiatkowski2019natural,yang2018hotpotqa}. Calibration uses 50,000 search states, one per question, sampled from trajectories generated on the training splits of five datasets (Appendix~\ref{Appendix:CalibrationSet}).

\paragraph{\upshape\bfseries Evaluation datasets}
Relative to the policy-optimization data, NQ~\cite{kwiatkowski2019natural} and HotpotQA~\cite{yang2018hotpotqa} provide in-domain evaluation, while PopQA~\cite{mallen2023trust}, TriviaQA~\cite{joshi2017triviaqa}, 2WikiMultiHopQA~\cite{ho2020constructing}, Bamboogle~\cite{press2023measuring}, and MuSiQue~\cite{trivedi2022musique} provide cross-dataset evaluation. NQ, TriviaQA, and PopQA form the single-hop group; the remaining four benchmarks form the multi-hop group. Evaluation splits and sample counts are listed in Appendix~\ref{Appendix:DatasetSplits}.

\paragraph{\upshape\bfseries Evaluation metrics}
Following prior agentic RAG work~\cite{song2025r1searcher,wu2026hiprag}, we measure answer accuracy with Cover Exact Match (CEM) and search efficiency with the average number of search calls per question. CEM scores an answer as correct when the ground-truth answer string appears in the extracted final answer, accommodating explanatory responses from instruction-tuned models.

\subsection{Baseline Definitions and Implementations}
\label{sec:baselines}
We compare against baselines spanning non-agentic retrieval, prompt-based agentic RAG, and reinforcement-learning-based agentic search. In our controlled evaluation, all baselines are implemented in the same codebase following their published algorithms and are evaluated under the shared settings in Appendix~\ref{Appendix:BackboneRetrievalSettings}.
All methods in Table~\ref{tab:hiprag-original-results} use the Qwen2.5-7B backbone.

\paragraph{\upshape\bfseries Non-agentic baselines}
\textit{Direct Inference} generates an answer from parametric knowledge without retrieval. \textit{Standard RAG} performs a single retrieval using the original question and conditions generation on the returned passages~\cite{lewis2020retrieval}.

\paragraph{\upshape\bfseries Prompt-based agentic RAG}
\textit{IRCoT} interleaves chain-of-thought reasoning and retrieval through prompting~\cite{trivedi2023ircot}. \textit{Search-o1} uses a prompt-driven multi-step search-and-reasoning procedure~\cite{li2025searcho1}.

\paragraph{\upshape\bfseries RL-based agentic RAG}
\textit{Search-R1} and \textit{R1-Searcher} learn interleaved reasoning and retrieval primarily from trajectory-level outcome supervision~\cite{jin2025searchr1,song2025r1searcher}. \textit{R1-Searcher++} adds cold-start supervised fine-tuning and rewards for internal-knowledge use~\cite{song2025r1searcherplus}. \textit{Search Wisely} introduces the $\beta$-GRPO objective to regulate search using model confidence~\cite{wu-etal-2025-search-wisely}. \textit{HiPRAG} provides hierarchical, knowledge-grounded process rewards over parsed search and no-search steps to encourage adaptive retrieval~\cite{wu2026hiprag}. In our controlled evaluation, all RL-based methods use the same policy-optimization data, number of training steps, rollout budget, and decoding constraints.

\paragraph{\upshape\bfseries Implementation settings}
We use Qwen2.5-7B-Instruct~\cite{qwen2024qwen25} as the main backbone, with Qwen2.5-3B-Instruct and Llama-3.2-3B-Instruct~\cite{meta2024llama32} for supplementary evaluation. Each search retrieves the top three passages from a shared Wikipedia index using E5-base~\cite{wang2022e5}, with at most four search turns per trajectory. Training runs for 400 optimization steps with five rollouts per prompt. We optimize all policy parameters with a learning rate of $5\times10^{-7}$. We use $K_{\mathrm{cal}}=32$ and $\lambda_f=\lambda_p=0.2$. Appendices~\ref{Appendix:BackboneRetrievalSettings} and~\ref{Appendix:Hyperparameters} provide the remaining settings.

\section{Experimental Results}
We first evaluate answer quality and search allocation across datasets and backbones, then isolate the contributions of the two scores and their reward weight. Behavioral and external-judge analyses examine what the scores capture at individual decisions. Finally, we report the training cost of this supervision.
\subsection{Main Results}
\label{sec:main-results}
We evaluate answer quality across seven QA benchmarks and search efficiency in a controlled 7B comparison.

\noindent\textbf{Best overall CEM.} As reported in Table~\ref{tab:hiprag-original-results}, our 7B model achieves 69.3 average CEM, exceeding HiPRAG, the strongest baseline, by 2.1 points. Improvements over HiPRAG cover all seven benchmarks, with multi-hop gains ranging from 0.1 point on 2Wiki to 5.6 points on Bamboogle, as shown in Figure~\ref{fig:adaptive-search-allocation}b. Gains on all five cross-dataset benchmarks suggest transfer beyond the datasets used for policy optimization.

\noindent\textbf{Controlled comparison.} Under the shared 7B setup (Section~\ref{sec:baselines}), search counts are averaged over all evaluation questions, including those answered without retrieval. The aggregate results in Table~\ref{tab:controlled-comparison} in Appendix~\ref{Appendix:CompleteSearchResults} therefore weight each dataset by its evaluation size.

\noindent\textbf{6.8\% fewer searches in the controlled 7B evaluation.} Table~\ref{tab:controlled-comparison} reports approximately 1.64 search calls per question for our policy, compared with 1.76 for our HiPRAG reimplementation and 1.83 for Search Wisely. The reduction relative to Search Wisely is 10.4\%.

\noindent\textbf{Fewer searches on single-hop QA, more on multi-hop QA in aggregate.} As shown in Table~\ref{tab:search-calls-selected}, our controlled 7B policy uses approximately 20.9\% fewer searches on single-hop QA and 6.6\% more on multi-hop QA than HiPRAG. This direction holds for every dataset: search decreases on all three single-hop benchmarks and increases on all four multi-hop benchmarks. Complete dataset-level search counts are given in Appendix~\ref{Appendix:CompleteSearchResults}. The consistent directions within each group indicate a task-dependent allocation of retrieval toward multi-step reasoning.

\noindent\textbf{Gains extend across model scales and families.} Table~\ref{tab:backbone-generalization} shows that our method exceeds HiPRAG by 1.5 CEM points on Qwen2.5-3B, 2.1 on Qwen2.5-7B, and 1.3 on Llama-3.2-3B. The same search-control formulation yields gains across the evaluated backbones when paired with a model-specific reliability mapping. These consistent gains support the generalizability of our method across model scales and families.

\begin{table}[!t]
\centering
\caption{Average CEM across backbones.}
\label{tab:backbone-generalization}
\begingroup
\fontsize{9}{10.8}\selectfont
\setlength{\tabcolsep}{3.2pt}
\renewcommand{\arraystretch}{1.08}
\setlength{\heavyrulewidth}{0.7pt}
\setlength{\lightrulewidth}{0.35pt}
\arrayrulecolor{black!65}
\begin{tabularx}{\columnwidth}{@{}Xr@{}}
\toprule
\textbf{Method} & \textbf{Avg. CEM} $\uparrow$ \\
\midrule
\rowcolor{blue!5}
\multicolumn{2}{l}{\rule[-2pt]{0pt}{12pt}\textbf{Qwen2.5-3B}} \\
HiPRAG & 64.4 \\
Ours & \textbf{65.9} \\
\addlinespace[2pt]
\rowcolor{blue!5}
\multicolumn{2}{l}{\rule[-2pt]{0pt}{12pt}\textbf{Qwen2.5-7B-Instruct}} \\
HiPRAG & 67.2\\
Ours & \textbf{69.3} \\
\addlinespace[2pt]
\rowcolor{blue!5}
\multicolumn{2}{l}{\rule[-2pt]{0pt}{12pt}\textbf{Meta-Llama-3.2-3B-Instruct}} \\
HiPRAG & 64.8\\
Ours & \textbf{66.1} \\
\bottomrule
\end{tabularx}
\endgroup
\par\vspace{10pt}
\begingroup
\renewenvironment{table}[1][]{}{}
\captionsetup{type=table}
\begin{table}[!b]
\centering
\caption{Ablation of reliability scores and mapping. Both uses the current-state score $S_t^{\mathrm{cur}}$ and the search-action score $\Delta R_t$; $f_m$ denotes the empirical reliability mapping.}
\label{tab:reward-ablation}
\begingroup
\fontsize{9}{10.8}\selectfont
\setlength{\tabcolsep}{3pt}
\renewcommand{\arraystretch}{1.12}
\begin{tabularx}{\columnwidth}{@{}>{\raggedright\arraybackslash}Xccr@{}}
\toprule
\textbf{Variant} & Scores & Scale & CEM $\uparrow$ \\
\midrule
Base & -- & -- & 61.2 \\
Current-state only & $S_t^{\mathrm{cur}}$ & $f_m$ & 63.4 \\
Search-action only & $\Delta R_t$ & $f_m$ & 62.3 \\
Raw confidence & Both & $\exp(-U)$ & 61.9 \\
\midrule
\textbf{Full} & \textbf{Both} & $\boldsymbol{f_m}$ & \textbf{65.9} \\
\bottomrule
\end{tabularx}
\endgroup
\end{table}
\endgroup

\par\vspace{10pt}
\captionsetup{type=figure}
\centering
\includegraphics[width=\linewidth]{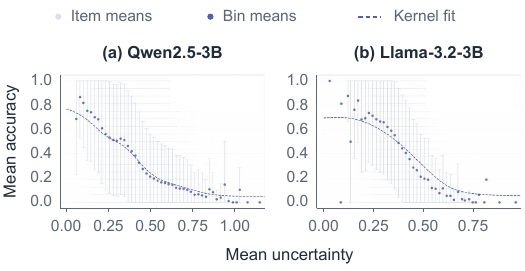}
\caption{Local uncertainty--accuracy relationships for (a) Qwen2.5-3B and (b) Llama-3.2-3B, with 32 conclusions per state. Bars indicate within-bin SD.}
\label{fig:additional-uncertainty-accuracy}
\end{table}

\begin{table}[!t]
\begingroup
\renewenvironment{table}[1][]{}{}
\captionsetup{type=table}
\begin{table}[!t]
\centering
\caption{Effect of the process-reward weight with Qwen2.5-3B-Instruct.}
\label{tab:process-reward-parameter-analysis}
\begingroup
\fontsize{9}{10.8}\selectfont
\setlength{\tabcolsep}{4pt}
\renewcommand{\arraystretch}{1.08}
\begin{tabularx}{\columnwidth}{@{}>{\raggedright\arraybackslash}Xr@{}}
\toprule
\textbf{Process weight $\lambda_p$} & CEM $\uparrow$ \\
\midrule
0 & 61.2 \\
0.1 & 63.8 \\
\textbf{0.2 (default)} & \textbf{65.9} \\
0.4 & 64.5 \\
\bottomrule
\end{tabularx}
\endgroup
\end{table}
\endgroup
\par\vspace{10pt}
\captionsetup{type=figure}
\centering
\includegraphics[width=\linewidth]{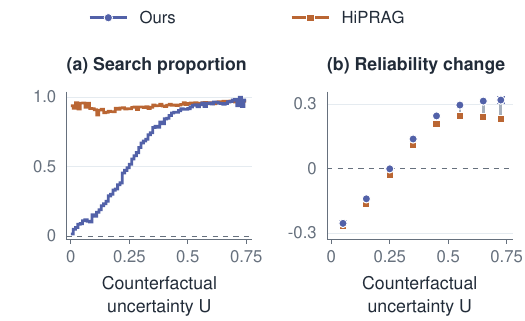}
\caption{Search behavior by no-evidence uncertainty. (a) Search proportion. (b) Reliability change $\Delta R$; connectors compare method means and whiskers mark uncertainty intervals.}
\label{fig:reliability-aware-search-behavior}
\end{table}

\subsection{Ablation Study}
\label{sec:ablation}

We ablate the two action-level scores and the empirical mapping. \textit{Base} retains only the answer and format rewards ($\lambda_p=0$). The single-score variants retain only $S_t^{\mathrm{cur}}$ or $\Delta R_t$, while \textit{Full} aggregates both according to Eq.~\ref{eq:process-reward}.

\noindent\textbf{Complementary roles of the two reliability scores.} As shown in Table~\ref{tab:reward-ablation}, combining the two scores yields 65.9 CEM, compared with 63.4 for current-state scoring and 62.3 for search-action scoring. The advantage of combining both scores suggests that search and no-search supervision are complementary. The single-score variants also improve over Base (61.2) by 2.2 and 1.1 points, respectively.

\noindent\textbf{Effect of the empirical reliability mapping.} Table~\ref{tab:reward-ablation} shows that, with probe contexts and process weight held fixed, Full reaches 65.9 CEM, compared with 61.9 for \textit{Raw confidence}. This 4.0-point gain favors the empirical uncertainty--accuracy mapping over raw confidence for process supervision. The Raw confidence control uses the same probe protocol and replaces the mapping with \mbox{$C(U)=\exp(-U)$}:
\begin{equation}
\begin{aligned}
S_{t,\mathrm{raw}}^{\mathrm{cur}}
&=C(\overline{U}_t^{\mathrm{ns}})-C(\mu_m^{\mathrm{cal}}),\\
\Delta C_t
&=C(\overline{U}_t^{\mathrm{evid}})-C(\overline{U}_t^{\mathrm{cf}}).
\end{aligned}
\label{eq:raw-confidence}
\end{equation}

Here, $\overline{U}_t^z$ is the mean probe uncertainty under condition $z$, and $\mu_m^{\mathrm{cal}}$ is the mean uncertainty across calibration states.

\noindent\textbf{Sensitivity to the process-reward weight.} As shown in Table~\ref{tab:process-reward-parameter-analysis}, using the default $\lambda_p=0.2$ raises CEM from 61.2 to 65.9 relative to $\lambda_p=0$. All tested nonzero weights improve over the baseline, with 0.2 performing best. Weights of 0.1 and 0.4 yield gains of 2.6 and 3.3 points, respectively; increasing the weight from 0.2 to 0.4 reduces CEM by 1.4 points. Only the process-reward weight varies; the answer and format rewards remain unchanged (Eq.~\ref{eq:total-reward}).

\begin{figure}[!t]
\centering
\includegraphics[width=\linewidth]{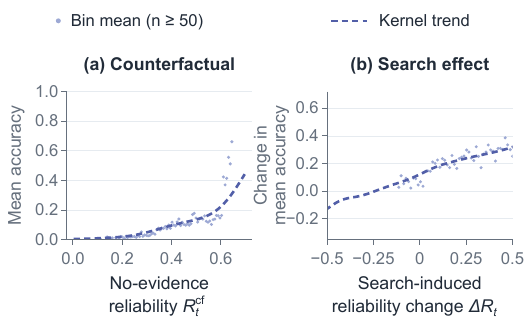}
\caption{External judgments across 10,000 paired search states. (a) No-evidence reliability and judged accuracy. (b) Their paired changes with evidence.}
\label{fig:continuous-reliability-validation}
\par\vspace{10pt}
\captionsetup{type=figure}
\centering
\includegraphics[width=\linewidth]{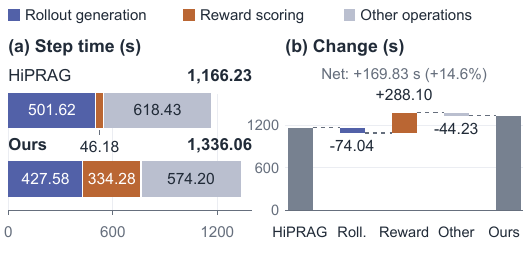}
\caption{Mean optimization-step time: (a) component times; (b) changes relative to HiPRAG.}
\label{fig:training-time-overhead}
\Description{Component times and their changes for HiPRAG and our method, with a net increase of 169.83 seconds per optimization step.}
\end{figure}

\begin{figure*}[!t]
\centering
\includegraphics[width=0.85\textwidth]{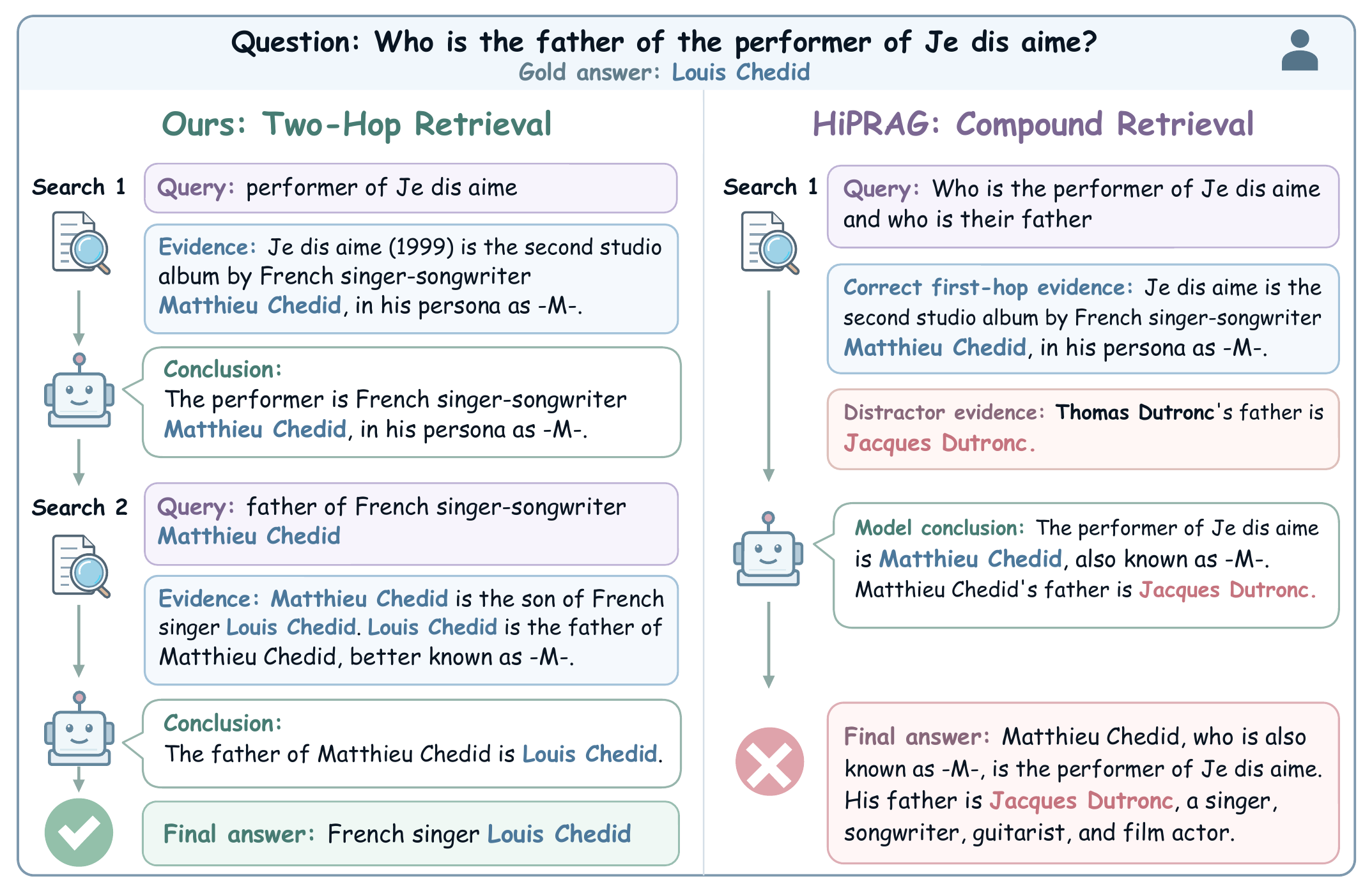}
\caption{MuSiQue case study: Ours (left) verifies the father relation with a second search; HiPRAG (right) identifies the performer but assigns a distractor's father to him.}
\label{fig:case-study}
\label{fig:case-study-ours}
\label{fig:case-study-hiprag}
\Description{For the question about the father of the performer of Je dis aime, our two-search trajectory identifies Matthieu Chedid and verifies his father, Louis Chedid. HiPRAG uses one compound query, retrieves Matthieu Chedid alongside distractor Thomas Dutronc, and incorrectly answers Jacques Dutronc.}
\end{figure*}

\subsection{Reliability-Aware Search Behavior}
\label{sec:search-behavior}
Dataset-level search counts do not reveal whether the policy responds to uncertainty at individual decisions. We therefore examine search frequency and evidence-induced reliability changes across local uncertainty levels in Figure~\ref{fig:reliability-aware-search-behavior}.

\noindent\textbf{Search frequency increases with uncertainty.} As shown in Figure~\ref{fig:reliability-aware-search-behavior}a, our policy searches less often than HiPRAG at low no-evidence counterfactual uncertainty, while both search frequently at high uncertainty. Figure~\ref{fig:reliability-aware-search-behavior}b shows larger mapped reliability gains $\Delta R$ for our selected searches than for HiPRAG's, especially at higher uncertainty. Table~\ref{tab:search-calls-selected} shows that, at the dataset level, search decreases on single-hop QA and increases on multi-hop QA in aggregate.

\subsection{Validating Reliability Estimates}
\label{sec:reliability-validation}

\noindent\textbf{Local accuracy decreases with mean conclusion uncertainty across backbones} under the same 32-conclusion protocol, as shown in Figures~\ref{fig:qwen7b-reliability} and~\ref{fig:additional-uncertainty-accuracy}. Differences in curve shape and uncertainty range motivate a separate mapping for each backbone.

Because $f_m$ is nonlinear, equal reductions in sequence uncertainty can produce different reliability gains. The search score depends on the uncertainty levels of both matched conditions and their difference. A given pair of uncertainty values can also yield different scores across backbones, whose empirical mappings differ in shape and range, as illustrated in Figures~\ref{fig:qwen7b-reliability} and~\ref{fig:additional-uncertainty-accuracy}.

\noindent\textbf{Higher no-evidence reliability $R^{\mathrm{cf}}$ is associated with higher judged accuracy} at search decisions sampled from the final policy, as shown in Figure~\ref{fig:continuous-reliability-validation}a. The judge compares each candidate with a reference conclusion for the same local step, using the question and decision context. References are supplied only to the judge. We average its binary correctness labels, not its confidence scores, to estimate local-answer accuracy.

\noindent\textbf{Larger reliability changes $\Delta R$ are associated with larger judged-accuracy differences} between evidence-conditioned and no-evidence conclusions, as shown in Figure~\ref{fig:continuous-reliability-validation}b. The paired conclusions share the same pre-retrieval state and query, so their judged-accuracy difference measures the change associated with the retrieved evidence. We summarize 10,000 paired search states using binned means and kernel-smoothed trends. The judge prompt and decoding settings are provided in Appendix~\ref{Appendix:ExternalJudgeValidation}.

\subsection{Efficiency Analysis}
\label{sec:practical-considerations}
\noindent\textbf{14.6\% more time per training step.} Figure~\ref{fig:training-time-overhead} shows that, relative to our HiPRAG reimplementation, additional reward-scoring time is partly offset by less time spent on rollout generation and other operations. Reward scoring takes longer because local probing requires sampling additional conclusions and computing their sequence uncertainty.

\FloatBarrier
\subsection{Case Study}
\label{sec:case-study}
\noindent\textbf{A targeted second search verifies the remaining relation.} MuSiQue example \texttt{dev\_98} requires the chain \textit{Je dis aime} $\rightarrow$ Matthieu Chedid (\textit{-M-}) $\rightarrow$ Louis Chedid. As shown on the left of Figure~\ref{fig:case-study}, our model uses two searches to identify the performer and verify his father. On the right of Figure~\ref{fig:case-study}, our HiPRAG reimplementation~\cite{wu2026hiprag} issues one compound query, retrieves the correct performer alongside distractor Thomas Dutronc, and incorrectly attributes the distractor's father, Jacques Dutronc, to the performer. The comparison illustrates that evidence sufficient to identify an entity may still leave the next relation unsupported.
\par

\begin{samepage}
\section{Conclusion}
We introduced a metacognitive framework for supervising search decisions through local-conclusion reliability. Across seven QA benchmarks, our Qwen2.5-7B model achieves 69.3 average CEM, exceeding HiPRAG by 2.1 points. In the controlled 7B evaluation, our policy uses 6.8\% fewer searches than our HiPRAG reimplementation, with fewer searches on single-hop tasks and more on multi-hop tasks in aggregate. Ablations support the complementary action scores and the empirical reliability mapping.
\par\end{samepage}

\clearpage
\appendix
\section{Additional Experimental Settings}
\label{Appendix:ExperimentalSettings}
\label{appendix:start}

\subsection{Dataset Splits and Sample Counts}
\label{Appendix:DatasetSplits}
The evaluation uses the dataset-specific development or test splits listed in Table~\ref{tab:evaluation-splits}. Following the grouping used in the main paper, NQ, PopQA, and TriviaQA are treated as single-hop QA benchmarks, whereas 2WikiMultiHopQA, Bamboogle, HotpotQA, and MuSiQue are treated as multi-hop QA benchmarks~\cite{kwiatkowski2019natural,mallen2023trust,joshi2017triviaqa,ho2020constructing,press2023measuring,yang2018hotpotqa,trivedi2022musique}. The complete evaluation set contains 51,713 examples.

\begin{table}[!t]
\centering
\caption{Evaluation splits and sample counts. Datasets are grouped following the main paper.}
\label{tab:evaluation-splits}
\small
\renewcommand{\arraystretch}{1.08}
\begin{tabular*}{\columnwidth}{@{\extracolsep{\fill}}lcr@{}}
\toprule
\textbf{Dataset} & \textbf{Split used} & \textbf{Samples}
\\
\midrule
\rowcolor{black!12}
\multicolumn{3}{l}{\textbf{Single-Hop QA}} \\
NQ & test & 3,610 \\
PopQA & test & 14,267 \\
TriviaQA & test & 11,313 \\
\midrule
\rowcolor{black!12}
\multicolumn{3}{l}{\textbf{Multi-Hop QA}} \\
2WikiMultiHopQA & dev & 12,576 \\
Bamboogle & test & 125 \\
HotpotQA & dev & 7,405 \\
MuSiQue & dev & 2,417 \\
\midrule
\multicolumn{2}{l}{\textbf{Total}} & \textbf{51,713} \\
\bottomrule
\end{tabular*}
\end{table}

\subsection{Calibration-Set Construction}
\label{Appendix:CalibrationSet}
We sampled 50,000 search states, one per question, from trajectories on the official training splits of NQ, TriviaQA, HotpotQA, 2WikiMultiHopQA, and MuSiQue.

\subsection{Backbone, Retrieval, and Evaluation Configuration}
\label{Appendix:BackboneRetrievalSettings}
We use Qwen2.5-7B-Instruct~\cite{qwen2024qwen25} as the main backbone, with Qwen2.5-3B-Instruct and Llama-3.2-3B-Instruct~\cite{meta2024llama32} for supplementary evaluation. All methods in our controlled 7B evaluation use the same E5-base retriever~\cite{wang2022e5} over the same indexed Wikipedia dump. Each search call returns the top three passages, and all methods share the retrieval corpus, maximum search budget, generation limits, and decoding protocol.

\subsection{Optimization and Decoding Hyperparameters}
\label{Appendix:Hyperparameters}
We weight the answer, format, and process terms by $0.8$, $0.2$, and $0.2$, respectively; equivalently, $\lambda_f=0.2$ and $\lambda_p=0.2$. The process term is added independently of the answer and format rewards (Eq.~\ref{eq:total-reward}). The remaining principal training and decoding hyperparameters are summarized in Table~\ref{tab:training-hyperparameters}.

\begin{table}[!t]
\centering
\caption{Principal training and decoding hyperparameters.}
\label{tab:training-hyperparameters}
\small
\renewcommand{\arraystretch}{1.08}
\begin{tabular*}{\columnwidth}{@{\extracolsep{\fill}}lr@{}}
\toprule
\textbf{Hyperparameter} & \textbf{Value} \\
\midrule
Training steps & 400 \\
GPUs & 4 \\
Train batch size & 512 \\
Rollouts per prompt & 5 \\
RL learning rate & $5\times10^{-7}$ \\
RL parameter updates & Full-parameter training \\
Policy mini-batch size & 256 \\
Policy micro-batch size per GPU & 2 (3B), 1 (7B) \\
Temperature & 1.0 \\
Maximum search turns & 4 \\
Retriever top-$k$ & 3 \\
\bottomrule
\end{tabular*}
\end{table}

\subsection{Calibration Context and Interaction Prompt}
\label{Appendix:Prompts}
We specify the calibration context and the structured prompt used for training and evaluation.

\subsubsection{Local Calibration Context}
Figure~\ref{fig:reliability-mapping-calibration-prompt} shows the calibration context: question, history, reasoning, and query, without the current evidence or original conclusion. References are used only for assessment.

\begin{figure}[!t]
    \centering
    \setlength{\fboxsep}{4pt}
    \colorbox{black}{\parbox{\dimexpr\columnwidth-2\fboxsep\relax}{\color{white}\fontsize{8.5}{10}\selectfont\bfseries Context for Local Reliability Calibration
        }}\\[-0.4pt]
    \fcolorbox{black}{gray!10}{\parbox{\dimexpr\columnwidth-2\fboxsep-2\fboxrule\relax}{\raggedright\fontsize{7.5}{9}\selectfont\ttfamily
            \textbf{Question:} \{question\}\\
            \textbf{History:} \{preceding trajectory\}\\
            \textbf{Reasoning:} \{current reasoning\}\\
            \textbf{Query:} \{current search query\}\\[5pt]
            \textbf{Sample:} 32 local conclusions
        }}
    \caption{Local calibration context. The current retrieval evidence and original conclusion are removed before sampling.}
    \label{fig:reliability-mapping-calibration-prompt}
\end{figure}

\subsubsection{Training and Evaluation Prompt}
We use the following structured prompt for both training and evaluation. It requires the model to organize its reasoning, optional search actions, retrieved context, and final answer using a consistent XML-style format.

\begin{figure}[!t]
    \centering
    \setlength{\fboxsep}{4pt}
    \colorbox{black}{\parbox{\dimexpr\columnwidth-2\fboxsep\relax}{\color{white}\fontsize{8.5}{10}\selectfont\bfseries Prompt for Training and Evaluation
        }}\\[-0.4pt]
    \begin{minipage}{\dimexpr\columnwidth-2\fboxsep-2\fboxrule\relax}
    \begin{lstlisting}[
        numbers=none,
        emptylines=0,
        frame=single,
        rulecolor=\color{black},
        backgroundcolor=\color{gray!10},
        basicstyle=\fontsize{7.5}{9}\selectfont\ttfamily,
        columns=fullflexible,
        keepspaces=true,
        breakindent=0pt,
        xleftmargin=0pt,
        framesep=4pt,
        framexleftmargin=0pt,
        framexrightmargin=0pt,
        framextopmargin=4pt,
        framexbottommargin=4pt,
        aboveskip=0pt,
        belowskip=0pt,
        breaklines=true,
        showstringspaces=false
]
Answer user questions by thinking step-by-step. Your entire reasoning process must be encapsulated within a single <think></think> block, which contains one or more <step></step> blocks. Each step must begin with your analysis in <reasoning>. If you identify a knowledge gap, you may use <search>query</search> to query a search engine; search results will then be provided in a <context> tag. Every step must end with a <conclusion> summarizing what you learned in that step. After your thinking process is complete, provide the final, conclusive answer inside an <answer> tag placed immediately after the closing </think> tag. Ensure all XML tags are properly formed and nested.

**## Output Format Specification**

<think>
<step>
    ...
</step>
<step>
    ...
</step>
</think>
<answer>Your final, conclusive answer to the user's question.</answer>

**## Step Formats (to be used inside <think>)**

Format 1: Step with a Search

<step>
    <reasoning>Your detailed analysis of what you know and what you need to find out.</reasoning>
    <search>The precise search query you will use.</search>
    <context>[This will be provided by the system after your search]</context>
    <conclusion>Your conclusion or answer to the reasoning and search query at this step.</conclusion>
</step>

Format 2: Step without a Search

<step>
    <reasoning>Your detailed analysis of what you know and what you need to find out.</reasoning>
    <conclusion>Your conclusion or answer to the reasoning at this step.</conclusion>
</step>

User Question: {question}
    \end{lstlisting}
    \end{minipage}
    \caption{Structured prompt used for trajectory generation during training and for model evaluation.}
    \label{fig:training-evaluation-prompt}
\end{figure}

\section{Supplementary Method and Analyses}
\label{Appendix:AdditionalResults}

\subsection{Construction of the Empirical Reliability Mapping}
\label{Appendix:ReliabilityMappingConstruction}

We fit one local uncertainty--accuracy mapping per backbone before optimization. The mapping, calibration mean, and reference $b_m$ remain fixed throughout the 400-step training run.

\paragraph{Local conclusions and uncertainty estimation.}
For each calibration state $i$ (Section~\ref{Appendix:CalibrationSet}), the frozen backbone independently generates $K_{\mathrm{cal}}=32$ local conclusions $y_{i,k}=(z_{i,k,1},\ldots,z_{i,k,T_{i,k}})$ from context $c_i$. We compute each conclusion's mean generated-token negative log-likelihood, excluding padding:
\begin{equation}
U_{i,k}
=-\frac{1}{T_{i,k}}\sum_{t=1}^{T_{i,k}}
\log p_m(z_{i,k,t}\mid c_i,z_{i,k,<t}).
\end{equation}
Qwen3.7-Flash~\cite{alibaba2026qwen37flash} checks the local reference against the dataset's human annotations, then judges whether each conclusion fully answers the local objective. Let $A_{i,k}\in\{0,1\}$ be this correctness label. Each state contributes the pair
\begin{equation}
\overline{U}_i=\frac{1}{K_{\mathrm{cal}}}\sum_{k=1}^{K_{\mathrm{cal}}}U_{i,k},
\qquad
\overline{A}_i=\frac{1}{K_{\mathrm{cal}}}\sum_{k=1}^{K_{\mathrm{cal}}}A_{i,k},
\end{equation}
where $\overline{U}_i$ is mean conclusion uncertainty and $\overline{A}_i$ is the fraction of conclusions judged correct.

\paragraph{Uncertainty binning and empirical centroids.}
We divide the observed range of $\overline{U}_i$ into 100 equal-width intervals $\{B_j\}_{j=1}^{100}$. For each nonempty interval, let $n_j=|B_j|$ and compute the centroid
\begin{equation}
(\Phi_j,\Psi_j)
=
\left(
\frac{1}{n_j}\sum_{i\in B_j}\overline{U}_i,
\frac{1}{n_j}\sum_{i\in B_j}\overline{A}_i
\right).
\end{equation}
Here, $\Phi_j$ and $\Psi_j$ are the interval's mean uncertainty and empirical accuracy. We retain the index set $\mathcal{J}_{50}=\{j:n_j\geq 50\}$ for both curve estimation and visualization.

\paragraph{Continuous reliability mapping.}
We use the count-weighted Gaussian estimator in Eq.~\ref{eq:empirical-mapping}, with bandwidth $h=0.055$, separately for each initial backbone. With $\mathcal{D}_{\mathrm{cal}}$ indexing the calibration states, the reference is
\begin{equation}
\mu_m^{\mathrm{cal}} =
\frac{1}{|\mathcal{D}_{\mathrm{cal}}|}
\sum_{i\in\mathcal{D}_{\mathrm{cal}}}\overline{U}_i,
\qquad b_m=f_m(\mu_m^{\mathrm{cal}}).
\end{equation}
Throughout optimization, all three probe conditions use this fixed mapping, and the no-search score uses the reference $b_m$. Section~\ref{Appendix:ExternalJudgeValidation} analyzes no-evidence reliability and paired reliability changes against external judgments of local conclusions for sampled search decisions.

\paragraph{Local-conclusion probe aggregation.}
At step $t$, let $h_t$, $r_t$, $q_t$, $c_t$, and $k_t$ denote the preceding history, current reasoning, search query, retrieved evidence, and original local conclusion, respectively. We remove $k_t$ before constructing a probe. A no-search probe conditions on the preceding history and current reasoning, $(h_t,r_t)$. For a search decision, the evidence-conditioned probe uses $(h_t,r_t,q_t,c_t)$, while the no-evidence counterfactual probe uses $(h_t,r_t,q_t)$. Both search conditions retain the same query and exclude the original search-derived conclusion; only the retrieved context differs. For each applicable condition, the current policy samples new local conclusions. General decoding hyperparameters are listed in Table~\ref{tab:training-hyperparameters}. Let $\overline{U}_t^z$ denote the mean uncertainty of the sampled conclusions under condition $z$. Applying the fixed mapping gives
\begin{equation}
R_t^z=f_m(\overline{U}_t^z).
\end{equation}
Here, $z\in\{\mathrm{ns},\mathrm{evid},\mathrm{cf}\}$ indexes the no-search, evidence-conditioned, and no-evidence counterfactual conditions. Only the newly generated local-conclusion tokens contribute to each sequence-level uncertainty; the trajectory prefix and, when applicable, retrieved evidence form the conditioning context. No-search scores subtract the reference $b_m$, while search scores subtract the counterfactual reliability from the evidence-conditioned reliability. Their rollout mean contributes an additive process term to the total reward in Eq.~\ref{eq:total-reward}. These probes are not used at inference.

\begingroup
\begin{table*}[!t]
\centering
\caption{Complete mean search calls from our controlled evaluation, reported for every dataset and method. Calls are averaged per question. Overall denotes the mean weighted by evaluation sample counts across all 51,713 examples.}
\label{tab:search-calls-complete}
\begingroup
\fontsize{9}{10.8}\selectfont
\setlength{\tabcolsep}{3.2pt}
\renewcommand{\arraystretch}{1.12}
\setlength{\heavyrulewidth}{0.7pt}
\setlength{\lightrulewidth}{0.35pt}
\setlength{\cmidrulewidth}{0.3pt}
\arrayrulecolor{black!65}
\begin{tabularx}{\textwidth}{l*{8}{>{\raggedleft\arraybackslash}X}}
\toprule
& \multicolumn{3}{c}{Single-hop QA} & \multicolumn{4}{c}{Multi-hop QA} & \\
\cmidrule(lr){2-4}\cmidrule(lr){5-8}
Method & NQ & PopQA & TriviaQA & 2Wiki & Bamboogle & HotpotQA & MuSiQue & Overall \\
\midrule
\rowcolor{blue!5}
\multicolumn{9}{l}{\rule[-2pt]{0pt}{12pt}Non-agentic baselines} \\
Direct Inference & 0.00 & 0.00 & 0.00 & 0.00 & 0.00 & 0.00 & 0.00 & 0.00 \\
Standard RAG & 1.00 & 1.00 & 1.00 & 1.00 & 1.00 & 1.00 & 1.00 & 1.00 \\
\addlinespace[3pt]
\rowcolor{blue!5}
\multicolumn{9}{l}{\rule[-2pt]{0pt}{12pt}Prompt-based agentic RAG} \\
IRCoT & 1.41 & 1.25 & 1.30 & 2.06 & 1.70 & 1.70 & 2.18 & 1.58 \\
Search-o1 & 1.08 & 1.04 & 1.08 & 1.27 & 1.38 & 1.26 & 1.44 & 1.16 \\
\addlinespace[3pt]
\rowcolor{blue!5}
\multicolumn{9}{l}{\rule[-2pt]{0pt}{12pt}RL-based agentic RAG} \\
Search-R1 & 1.27 & 1.31 & 1.30 & 1.89 & 1.66 & 1.67 & 2.05 & 1.53 \\
R1-Searcher & 1.19 & 1.21 & 1.20 & 1.75 & 1.49 & 1.53 & 1.80 & 1.41 \\
Search Wisely & 1.36 & 1.65 & 1.55 & 2.26 & 2.02 & 1.94 & 2.36 & 1.83 \\
HiPRAG & 1.31 & 1.57 & 1.49 & 2.18 & 1.90 & 1.88 & 2.27 & 1.76 \\
\midrule
Ours & 1.09 & 1.21 & 1.20 & 2.38 & 1.94 & 1.91 & 2.43 & 1.64 \\
\bottomrule
\end{tabularx}
\endgroup
\end{table*}
\endgroup

\subsection{External-Judge Analysis of Reliability Estimates and Changes}
\label{Appendix:ExternalJudgeValidation}

We analyze 10,000 paired search states sampled from trajectories of the final policy. For each state, we construct matched evidence-conditioned and no-evidence counterfactual probes with the same pre-retrieval history, current reasoning, and query. Both exclude the original local conclusion; the counterfactual also omits the current retrieved evidence while retaining earlier evidence in the history. Both reliability estimates use the fixed mapping constructed before policy optimization. Candidate conclusions used as external labels are sampled independently from those used to compute the scores.

Independently sampled conclusions yield judged accuracies $A_t^{\mathrm{evid}}$ and $A_t^{\mathrm{cf}}$ under the matched evidence conditions. Higher no-evidence reliability $R_t^{\mathrm{cf}}$ is associated with higher $A_t^{\mathrm{cf}}$ in Figure~\ref{fig:continuous-reliability-validation}a. The paired accuracy change,
\begin{equation}
\Delta A_t=A_t^{\mathrm{evid}}-A_t^{\mathrm{cf}},
\end{equation}
increases with the reliability change $\Delta R_t=R_t^{\mathrm{evid}}-R_t^{\mathrm{cf}}$ in Figure~\ref{fig:continuous-reliability-validation}b.

We use GPT-5.6-luna~\cite{openai2026gpt56luna} as the external judge. It receives the original question, trajectory history and current reasoning, candidate conclusion, and a reference conclusion for the local step. Matched evidence conditions share the same reference, supplied only to the judge and excluded from probe generation and reliability scoring. Current retrieved passages are withheld. The judge checks whether the candidate resolves the local objective and agrees factually with the reference, allowing equivalent wording and aliases. Decoding uses temperature 0 and a maximum output length of 180 tokens. The returned JSON contains a binary \texttt{correct} label, a confidence score in $[0,1]$, and a brief reason. Only \texttt{correct} enters accuracy calculations; confidence is diagnostic. In Figure~\ref{fig:continuous-reliability-validation}, pale points show bin means, and dashed curves show kernel-smoothed trends. Each panel starts with 100 equal-width bins over its observed reliability range. Panel (a) displays bins containing at least 50 states; panel (b) merges adjacent bins until each contains at least 50 states, merging a final smaller remainder into the preceding bin.

\begin{figure}[!t]
    \centering
    \setlength{\fboxsep}{4pt}
    \colorbox{black}{\parbox{\dimexpr\columnwidth-2\fboxsep\relax}{\color{white}\fontsize{8.5}{10}\selectfont\bfseries External-Judge System Prompt
        }}\\[-0.4pt]
    \begin{minipage}{\dimexpr\columnwidth-2\fboxsep-2\fboxrule\relax}
    \begin{lstlisting}[
        numbers=none,
        emptylines=0,
        frame=single,
        rulecolor=\color{black},
        backgroundcolor=\color{gray!10},
        basicstyle=\fontsize{7.5}{9}\selectfont\ttfamily,
        columns=fullflexible,
        keepspaces=true,
        breakindent=0pt,
        xleftmargin=0pt,
        framesep=4pt,
        framexleftmargin=0pt,
        framexrightmargin=0pt,
        framextopmargin=4pt,
        framexbottommargin=4pt,
        aboveskip=0pt,
        belowskip=0pt,
        breaklines=true,
        showstringspaces=false
]
You are a strict factual-conclusion judge. Evaluate whether the candidate correctly resolves the local objective defined by the question and decision context, using the reference local conclusion as the factual standard. Accept equivalent wording, aliases, and minor spelling variants. Mark incorrect if the candidate contradicts the reference, gives an incompatible entity or relation, or fails to answer the local objective. Do not reward fluency, confidence, or verbosity. Return JSON only: {"correct": true or false, "confidence": number from 0 to 1, "reason": "brief reference-based reason"}. Confidence is diagnostic only.
    \end{lstlisting}
    \end{minipage}

    \colorbox{black}{\parbox{\dimexpr\columnwidth-2\fboxsep\relax}{\color{white}\fontsize{8.5}{10}\selectfont\bfseries External-Judge User Prompt Template
        }}\\[-0.4pt]
    \begin{minipage}{\dimexpr\columnwidth-2\fboxsep-2\fboxrule\relax}
    \begin{lstlisting}[
        numbers=none,
        emptylines=0,
        frame=single,
        rulecolor=\color{black},
        backgroundcolor=\color{gray!10},
        basicstyle=\fontsize{7.5}{9}\selectfont\ttfamily,
        columns=fullflexible,
        keepspaces=true,
        breakindent=0pt,
        xleftmargin=0pt,
        framesep=4pt,
        framexleftmargin=0pt,
        framexrightmargin=0pt,
        framextopmargin=4pt,
        framexbottommargin=4pt,
        aboveskip=0pt,
        belowskip=0pt,
        breaklines=true,
        showstringspaces=false
]
Original question: {original_question}
Decision context: {trajectory_history_and_current_reasoning}
Reference local conclusion: {reference_local_conclusion}
Candidate local conclusion: {candidate_local_conclusion}

Judge whether the candidate correctly resolves the local objective using the reference conclusion.
    \end{lstlisting}
    \end{minipage}
    \caption{System prompt and user-prompt template for reference-guided evaluation of independently generated local conclusions.}
    \label{fig:external-judge-prompt}
\end{figure}

\section{Controlled Comparison and Dataset-Level Search Results}
\label{Appendix:CompleteSearchResults}

Table~\ref{tab:search-calls-complete} reports all per-dataset search counts from our controlled 7B evaluation in Section~\ref{sec:main-results}. Overall denotes the mean weighted by evaluation sample counts across all 51,713 examples.

\begin{table}[!htbp]
\centering
\caption{Controlled comparison on seven QA benchmarks with Qwen2.5-7B-Instruct under shared training and retrieval settings.}
\label{tab:controlled-comparison}
\begingroup
\fontsize{9}{10.8}\selectfont
\setlength{\tabcolsep}{4pt}
\renewcommand{\arraystretch}{1.12}
\begin{tabularx}{\columnwidth}{@{}>{\raggedright\arraybackslash}Xrr@{}}
\toprule
Method & CEM $\uparrow$ & Search calls \\
\midrule
Search Wisely & 62.4 & 1.83 \\
HiPRAG & 67.2 & 1.76 \\
\textbf{Ours} & 69.3 & 1.64 \\
\bottomrule
\end{tabularx}
\endgroup
\end{table}
\end{document}